\documentclass[]{spie}  

\usepackage{amsmath,amsfonts,amssymb}
\usepackage{graphicx}
\usepackage{float}
\usepackage{hyperref}
\usepackage{booktabs}
\usepackage{multirow}

\title{Real-time topology-aware M-mode OCT segmentation for robotic deep anterior lamellar keratoplasty (DALK) guidance}

\author[a]{Rosalinda Xiong}
\author[a]{Jinglun Yu}
\author[a]{Yaning Wang}
\author[a]{Ziyi Huang}
\author[a]{Jin U.~Kang}

\affil[a]{Department of Electrical and Computer Engineering, Johns Hopkins University, Baltimore, MD 21218, USA}

\authorinfo{Further author information: (Send correspondence to Rosalinda Xiong)\\
Rosalinda Xiong: E-mail: rxiong5@jhu.edu}

\begin{document}
\maketitle

\begin{abstract}
Robotic deep anterior lamellar keratoplasty (DALK) requires accurate real-time depth feedback to approach Descemet's membrane (DM) without perforation.
M-mode intraoperative optical coherence tomography (OCT) provides high-temporal-resolution depth traces, but speckle noise, attenuation, and instrument-induced shadowing often result in discontinuous or ambiguous layer interfaces that challenge anatomically consistent segmentation at deployment frame rates.
We present a lightweight, topology-aware M-mode segmentation pipeline based on UNeXt that incorporates anatomical topology regularization to stabilize boundary continuity and layer ordering under low signal-to-noise ratio conditions.
The proposed system achieves end-to-end throughput exceeding 80~Hz measured over the complete preprocessing--inference--overlay pipeline on a single GPU, demonstrating practical real-time guidance beyond model-only timing.
This operating margin provides temporal headroom to reject low-quality or dropout frames while maintaining a stable effective depth update rate.
Evaluation on a standard rabbit-eye M-mode dataset using an established baseline protocol shows improved qualitative boundary stability compared with topology-agnostic controls, while preserving deployable real-time performance.
\end{abstract}


\section{INTRODUCTION}

Deep anterior lamellar keratoplasty (DALK) preserves Descemet's membrane (DM) and the corneal endothelium while replacing diseased stromal tissue, thereby reducing graft rejection risk compared to penetrating keratoplasty
\cite{wang2023common,wang2024optical,gensheimer2024comparison,wang2024reimagining}.
Safe advancement of a needle or cannula during DALK requires accurate depth awareness when operating near DM, where the risk of perforation is highest
\cite{opfermann2024novel,singh2024live,xu2023neural}.

Intraoperative OCT enables visualization of corneal microstructure during surgery.
M-mode OCT, in particular, provides high temporal resolution at a fixed lateral position, making it well suited for real-time depth monitoring and feedback.
However, M-mode images are strongly affected by speckle noise, attenuation, and instrument-induced shadowing.
These effects frequently lead to fragmented or ambiguous interface appearances, causing conventional segmentation networks to produce discontinuous boundaries or anatomically implausible outputs
\cite{yi2025kalman}.

Topology-aware loss formulations have been shown to improve structural coherence in medical image segmentation by enforcing global anatomical constraints
\cite{mirikharaji2018star}.
Separately, lightweight architectures such as UNeXt enable high-throughput inference suitable for real-time deployment
\cite{valanarasu2022unext}.
Building on these ideas, we introduce a topology-aware UNeXt-based M-mode segmentation pipeline designed for intraoperative depth guidance using solid-band overlays familiar to surgeons.
Performance is evaluated with an emphasis on deployability, reporting interface-level accuracy, segmentation-map fidelity (PSNR/SSIM), and end-to-end frame rate measured over the complete visualization pipeline
\cite{yu2025topology}.

\section{METHODS}

\subsection{Dataset and Preprocessing}

A rabbit-eye M-mode OCT dataset and corresponding labeling protocol widely used for DALK layer tracking are adopted in this study
\cite{yu2025topology,wang2024reimagining}.
The in vivo subset contains 500 annotated $512\times512$ M-mode images, while the ex vivo subset contains 250 annotated images of the same size.
A hybrid subset is constructed by merging the two.
Following the established protocol, train/test splits are set to 400/100 (in vivo), 200/50 (ex vivo), and 600/150 (hybrid), with an additional 20\% of the training portion reserved for validation.

Two anatomical interfaces are tracked: the upper epithelium boundary and the lower Descemet’s membrane boundary.
All distance-based metrics are reported using the dataset calibration of $2.61~\mu$m per pixel.

Each $512\times512$ M-mode frame is partitioned into eight non-overlapping vertical stripes of size $512\times64$.
Stripe intensities are standardized using statistics computed from the training set.
During inference, stripe-level predictions are reassembled into the original image for visualization and metric evaluation.

\subsection{Model Architecture and Inference}

UNeXt is employed with a single-channel input and $C$ output classes
\cite{valanarasu2022unext}.
To satisfy network downsampling constraints, inputs are zero-padded so that both spatial dimensions are multiples of 16, and output logits are cropped back to the original size:
\begin{equation}
x'=\mathrm{pad16}(x), \quad \hat{y}'=f_{\theta}(x'), \quad \hat{y}=\mathrm{crop}(\hat{y}',x).
\end{equation}
Automatic mixed precision (AMP) is used during inference, and stripe inputs are processed in batches to maximize throughput.

\subsection{End-to-End Deployment Pipeline}

Intraoperative usability is governed by the complete processing and visualization pipeline rather than isolated model inference.
Each incoming M-mode frame undergoes stripe tiling and normalization, host--device transfer, batched GPU inference, stripe reassembly, post-processing, and solid-band overlay rendering.
All reported frame rates correspond to this complete pipeline, reflecting the actual update rate delivered to the intraoperative display or robotic controller.

\subsection{Loss Function}

Training minimizes a weighted combination of region-based and topology-aware objectives, including
binary cross-entropy (BCE), Dice loss, and a topology-aware regularization term:
\begin{equation}
\mathcal{L} = \lambda_{\mathrm{CE}}\mathcal{L}_{\mathrm{CE}} +
\lambda_{\mathrm{Dice}}\mathcal{L}_{\mathrm{Dice}} +
\lambda_{\mathrm{topo}}\mathcal{L}_{\mathrm{topo}}.
\end{equation}
The topology term is implemented using a star-shape prior that discourages fragmented boundaries and anatomically implausible configurations along the depth axis
\cite{mirikharaji2018star,yu2025topology}.
The topology weight is gradually increased during training to avoid early over-regularization.

\subsection{Evaluation Metrics}

Segmentation performance is quantified using macro Dice/IoU, mean absolute boundary error (pixels and $\mu$m), SSIM, PSNR, and end-to-end frame rate.
PSNR is computed on normalized non-background segmentation maps with $\mathrm{MAX}=1$
\cite{wang2004ssim}.

\section{EXPERIMENTS AND RESULTS}

\subsection{Implementation Details}

All models are trained in PyTorch using the AdamW optimizer with AMP.
Data augmentation includes mild intensity jitter and horizontal flips.
Rotations are excluded to preserve M-mode acquisition geometry.
Model selection is based on validation boundary error (primary criterion) and macro Dice (secondary criterion).
All experiments are conducted on a single GPU.

\subsection{Prior Method Reference Performance}

Prior results reported in Yu \textit{et al.}~\cite{yu2025topology} are summarized in Table~\ref{tab:baseline_seg} and serve as strong literature baselines.
These methods use modified U-Net backbones with and without topology-aware regularization and are not architecturally equivalent to the proposed UNeXt-based approach.

\begin{table}[t]
\centering
\caption{Segmentation performance of prior method (PM) modified U-Net versus the proposed UNeXt method on the rabbit-eye M-mode OCT dataset. Pixel spacing: $2.61~\mu$m/px.}
\label{tab:baseline_seg}
\begin{tabular}{llccccc}
\toprule
Method & Data set & SSIM & PSNR & IoU & Dice & Hz \\
\midrule
PM-U-Net (topology-aware) & ex vivo & 0.9901 & 29.41 & 0.9874 & 0.9936 & 34 \\
PM-U-Net (BCE)           & ex vivo & 0.9845 & 24.96 & 0.9798 & 0.9898 & 25 \\
PM-U-Net (topology-aware) & hybrid & 0.9934 & 35.11 & 0.9908 & 0.9953 & 40 \\
PM-U-Net (BCE)           & hybrid & 0.9858 & 25.95 & 0.9813 & 0.9905 & 28 \\
U-NeXt (topology-aware)  & ex vivo & 0.9801 & 31.83 & 0.9332 & 0.9987 & 87 \\
U-NeXt (BCE)             & ex vivo & 0.9751 & 26.83 & 0.9897 & 0.9794 & 82 \\
U-NeXt (topology-aware)  & hybrid & 0.9847 & 36.21 & 0.9940 & 0.9892 & 84 \\
U-NeXt (BCE)             & hybrid & 0.9858 & 27.29 & 0.9901 & 0.9896 & 93 \\
\bottomrule
\end{tabular}
\end{table}

We note that in some cases PSNR improves while IoU exhibits a smaller gain.
This behavior reflects the different sensitivities of the metrics:
PSNR captures global pixel-wise fidelity of the predicted probability bands,
whereas IoU depends on hard region overlap after thresholding.
Topology-aware training promotes smoother and more temporally consistent layer bands,
which can improve PSNR and boundary stability while slightly altering discrete class boundaries
used for IoU computation.

\begin{table}[t]
\centering
\caption{Robustness performance analysis based on average absolute boundary tracking error
for the epithelium and Descemet's membrane (DM).
Lower values indicate more stable interface tracking under speckle, attenuation, and shadowing.
Pixel spacing: $2.61~\mu$m/px.}
\label{tab:robustness_boundary}
\setlength{\tabcolsep}{6pt}
\renewcommand{\arraystretch}{1.1}
\begin{tabular}{llcccc}
\toprule
Method & Data set & Epi (px) & Epi ($\mu$m) & DM (px) & DM ($\mu$m) \\
\midrule
PM-U-Net (topology-aware) & ex vivo & 0.34 & 0.8874 & 1.26 & 3.2886 \\
PM-U-Net (BCE)           & ex vivo & 0.75 & 1.9575 & 1.84 & 4.8024 \\
PM-U-Net (topology-aware) & hybrid & 0.38 & 0.9918 & 0.67 & 1.7487 \\
PM-U-Net (BCE)           & hybrid & 0.89 & 2.3229 & 1.41 & 3.6801 \\
U-NeXt (topology-aware)  & ex vivo & 0.41 & 0.7931 & 1.17 & 2.4526 \\
U-NeXt (BCE)             & ex vivo & 0.67 & 1.8055 & 1.71 & 4.2099 \\
U-NeXt (topology-aware)  & hybrid & 0.31 & 0.9796 & 0.55 & 1.6977 \\
U-NeXt (BCE)             & hybrid & 0.83 & 1.9985 & 1.39 & 3.7321 \\
\bottomrule
\end{tabular}
\end{table}

\begin{figure}[h]
\centering
\includegraphics[width=0.55\linewidth]{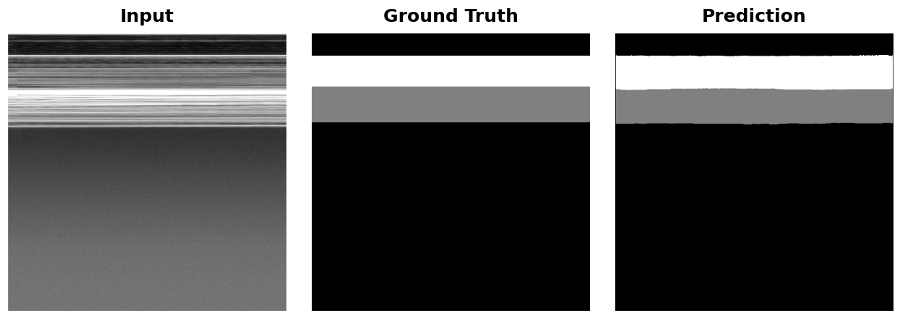}
\caption{Representative solid-band overlays produced by the proposed topology-aware UNeXt pipeline.
Topology regularization reduces boundary discontinuities and suppresses anatomically implausible layer crossings under low-SNR conditions.}
\label{fig:overlay}
\end{figure}

\subsection{Temporal Robustness and Dropout Tolerance}

Intraoperative M-mode streams frequently exhibit transient signal degradation due to speckle bursts, attenuation, or instrument shadowing.
Operating at an end-to-end rate exceeding typical display and control requirements provides temporal redundancy that can be exploited to reject low-quality frames while maintaining a stable effective update rate.
At a capped deployable operating point of 80~Hz, the proposed pipeline maintains sufficient headroom to support lightweight confidence-based gating without disrupting visual continuity.

\section{DISCUSSION AND CONCLUSION}

We presented a topology-aware UNeXt pipeline for real-time M-mode OCT segmentation to support robotic DALK depth guidance.
By enforcing anatomical consistency while preserving high-throughput inference, the proposed approach bridges the gap between segmentation accuracy and intraoperative deployability.
Compared with prior U-Net--based methods, topology-aware UNeXt enables stable solid-band depth overlays at substantially higher end-to-end frame rates.

Limitations include reliance on a single dataset and the inability of topology regularization to recover interfaces under complete signal loss.
Future work will focus on confidence-aware gating, closed-loop robotic safety constraints, and expanded validation across imaging systems and surgical scenarios.

\acknowledgments
This work was supported by the National Institute of Biomedical Imaging and Bioengineering of the National Institutes of Health under award number 1R01EY032127 (PI: Jin U.~Kang).

\bibliography{report}
\bibliographystyle{spiebib}

\end{document}